\documentclass[12pt,preprint]{aastex}

\slugcomment{accepted, Astrophysical Journal}

\shorttitle{Radio Bursts on AD Leo}
\shortauthors{Osten \& Bastian }

\begin{document}
 
\title{ Wideband Spectroscopy of Two Radio Bursts on AD~Leonis}
 
\author{Rachel A. Osten\altaffilmark{1}}
\author{T. S. Bastian}
\affil{National Radio Astronomy Observatory, 520 Edgemont Road, Charlottesville, VA 22903 \\
Electronic Mail: rosten@nrao.edu, tbastian@nrao.edu}
\altaffiltext{1}{Jansky Fellow, NRAO}

\begin{abstract}

We report high-time-resolution, broadband spectroscopic observations of two radio bursts on the 
classical flare star AD~Leonis.  The observations were acquired by the 305~m telescope at 
Arecibo Observatory on 2003 June 13-14. Using the Wideband Arecibo Pulsar Processor, these 
observations sampled a total bandwidth of 400 MHz, distributed over a 500 MHz frequency range, 
1120--1620 MHz, with a frequency resolution of 0.78 MHz and a time resolution of 10~ms. A radio 
burst observed on June 13 is characterized by the presence of multitudes of short duration ($\Delta 
t \sim$30 ms), high brightness temperature ($T_{b}>10^{14}$K), highly circularly polarized, fast-drift 
radio sub-bursts, with median bandwidths $\Delta \nu/\nu \sim$5\%.  The inverse drift rates are small, 
and have a symmetric distribution (both positive and negative frequency drifts) with a Gaussian FWHM 
inverse  drift rate of 4.5$\times$10$^{-4}$ s/MHz.  The fast-drift sub-bursts occur at a mean rate 
of 13 s$^{-1}$ and show no evidence for periodic recurrence. The  fast-drift radio events on AD~Leo 
are highly reminiscent of solar decimetric spike bursts. We suggest the emission is due to fundamental 
plasma radiation. A second highly circularly polarized radio burst, recorded June 14, has markedly 
different properties:  a smoothly varying intensity profile characterized by a slow drift in frequency 
with time ($-52$ MHz s$^{-1}$). Under the assumption that the source is due to a disturbance propagating 
through the low corona, a source size of 0.1--1 R$_{\star}$ is inferred, implying a brightness 
temperature range 6$\times$10$^{11}$--6$\times$10$^{13}$K: another example of a coherent radio burst.

\end{abstract}


\section{Introduction}

AD~Leonis, a young disk star at a distance from the Sun of 4.9 pc, is one of the most active 
flare stars known. AD~Leo produces intense, quasi-steady chromospheric and coronal emissions, 
with a 0.1--2.4 keV X-ray luminosity of 7$\times$10$^{28}$ erg s$^{-1}$ \citep{hunschetal1999}, 
C~IV $\lambda$ 1548 luminosity of 9.5$\times$10$^{26}$ erg s$^{-1}$ \cite{hawleyetal2003} and 
quiescent 20 cm radio luminosity of 5.5$\times$10$^{13}$ erg s$^{-1}$ Hz$^{-1}$ \citep{jackson1989}. 
The star is also highly variable, producing flares at radio, optical, UV, EUV, and X-ray wavelengths
 \citep[e.g.,][]{bastianetal1990,hawleypettersen1991,hawleyetal2003,hawleyetal1995,favataetal2000}.

Investigations of AD~Leo's coronal characteristics led \citet{gudeletal2003} to 
suggest that AD~Leo is constantly undergoing small-scale coronal flare events, which explains the 
low-level variability as well as the intriguing plasma characteristics: namely, the high coronal 
electron densities of several $\times$ 10$^{10}$ cm$^{-3}$ and a concentration of plasma at 
temperatures 6--20 MK. Coronal densities and temperatures of this order are typically seen on the 
Sun during flares but not in quiescence. The distribution of flare frequency with energy on AD~Leo 
is steep enough that flares may account for the entire observed X-ray luminosity, making the disparate 
nature of AD~Leo's corona compared with the Sun's corona even more striking.

Radio bursts on AD~Leo have been observed for many years, beginning with a report by \citet{spangler1974}.
  Radio bursts on AD~Leo are characterized by large flux density enhancements -- as large as $\approx$ 
500 times the quiescent flux density level of $\approx$ 2 mJy -- that can occur on very short time 
scales ($<0.02$ seconds). The degree of circular polarization of the radio bursts is typically very 
high, often reaching 100\% \citep{langetal1983, gudeletal1989, bastianetal1990, abada1997a}.  
The observations imply extremely large brightness temperatures during radio 
bursts, T$_{b} \geq$ 10$^{15}$K, and suggest that one or more coherent radiation mechanisms 
operate in the corona of AD~Leo. Both plasma radiation mechanisms and the cyclotron maser instability 
have been suggested as a means of accounting for the coherent radio emissions.  Neither has been 
definitively ruled out by the observations.

Spectroscopic investigations of the coherent radio bursts on flare stars have 
typically been hampered by relatively long integration times \citep{bb1987,gudeletal1989} and/or 
limited bandwidths \citep[e.g.,][]{bastianetal1990,abada1997a}. The necessary combination of high 
time resolution and broadband frequency coverage has only been available infrequently 
\citep{stepanov2001, zaitsev2004}, precluding the means to measure key parameters such as the 
intrinsic bandwidth of the radio bursts, thereby hampering their interpretation.  

The recent upgrade of the Arecibo Observatory 305~m telescope has yielded a number of important 
improvements that motivated us to revisit the problem of coherent radio bursts on AD~Leo. First, the 
installation of a ground screen mitigates the problem of beam spillover, enabling users to track 
targets for longer periods of time with high sensitivity. Second, due to more complete illumination 
of the 305 m primary by the Gregorian feed (illumination area 213 m$\times$237 m at zenith), the 
sensitivity of the telescope has improved. Finally, with the installation of Gregorian optics, a 
new broadband feed, and wideband receivers, the instantaneous bandwidth accessible has improved 
enormously. The relative bandwidth now accessible to Arecibo with the L-band wide receiver is 
$\Delta\nu/\nu=36\%$, ten times the relative bandwidth previously available at Arecibo 
\citep{bastianetal1990}, and more than three times the maximum relative bandwidths observed elsewhere 
\citep{stepanov2001, zaitsev2004}. We therefore initiated a pilot program to observe dMe flares stars 
with the Arecibo 305~m telescope in June 2003, resulting in the detection of two radio bursts.

\section{Observations \label{analsec}}

We observed AD~Leo from 2003 June 12--15, during which time approximately 
16 hours of data were collected. The observations were made using the ``L-band wide" dual-linear feed and 
receiver, which provides frequency coverage from 1.15--1.73 GHz, with the Wideband Arecibo 
Pulsar Processor (WAPP) backend.  The 
WAPP was selected because it provides the means of observing a large 
instantaneous bandwidth with excellent spectral and temporal resolution. The 
WAPP provides four data channels, each of 100 MHz bandwidth.  These were deployed across the L-band 
wide receiver as follows: 1120--1220 MHz, 1320--1420 MHz, 1420--1520 MHz, and 1520--1620 MHz.  A 
gap was deliberately left between 1220--1320 MHz to avoid the strong radio frequency
interference (RFI) present in this frequency range.  Nevertheless, the effects of other sources of 
RFI could not be entirely avoided in the frequency bands observed. We employed a data acquisition 
mode where data were sampled with a time resolution of 10 ms; 128 spectral channels were sampled 
across each 100 MHz channel, yielding a spectral resolution of 0.78 MHz. All four correlation products 
were recorded between the native-linear X and Y feed elements (XX, YY, XY, YX) with three-level sampling.  

The beam size using the L-band wide receiver at 1.4 GHz is $\approx3.1\times$3.5 arcminutes in 
azimuth and zenith angle. The 1$\sigma$ noise uncertainty ranges from 28--43 mJy per frequency and 
time bin for the different WAPPs. Interferometric measurements of AD~Leo's 1.4~GHz flux density 
during periods of apparent quiescence are $\approx2$ mJy \citep{jackson1989}, well below the confusion 
limit of Arecibo at the observed frequencies ($\approx$ 16--38 mJy from 1120 MHz to 1620 MHz). 
Our observing strategy was to observe the target, AD Leo, for 10 minute scans, and then to 
inject a correlated calibration signal to determine antenna temperatures. While there is a bright 
background radio source located $\approx$ 2.3 arcminutes away from AD~Leo \citep{seiradakis1995} and 
therefore within the primary beam, we utilized a ``time-switching'' scheme to difference times of 
burst activity (``on") 
and the quiescent state plus background (``off"). This procedure enabled us to 
estimate the antenna temperature and flux density through the use of gain curves as a function of 
azimuth and zenith angle. 

\section{Data Analysis and Results}

Two radio bursts were identified in the WAPP data. The first radio burst occurred from approximately 
22:35:20--23:36:30 UT on 2003 June 13 and the second 
occurred from approximately 20:05:40--20:06:30 UT on 
2003 June 14.  We now describe the observations, their analysis, and results for each event in turn.  


\subsection{Event on June 13}
 
\begin{figure}[h]
\begin{center}
\includegraphics[scale=0.6]{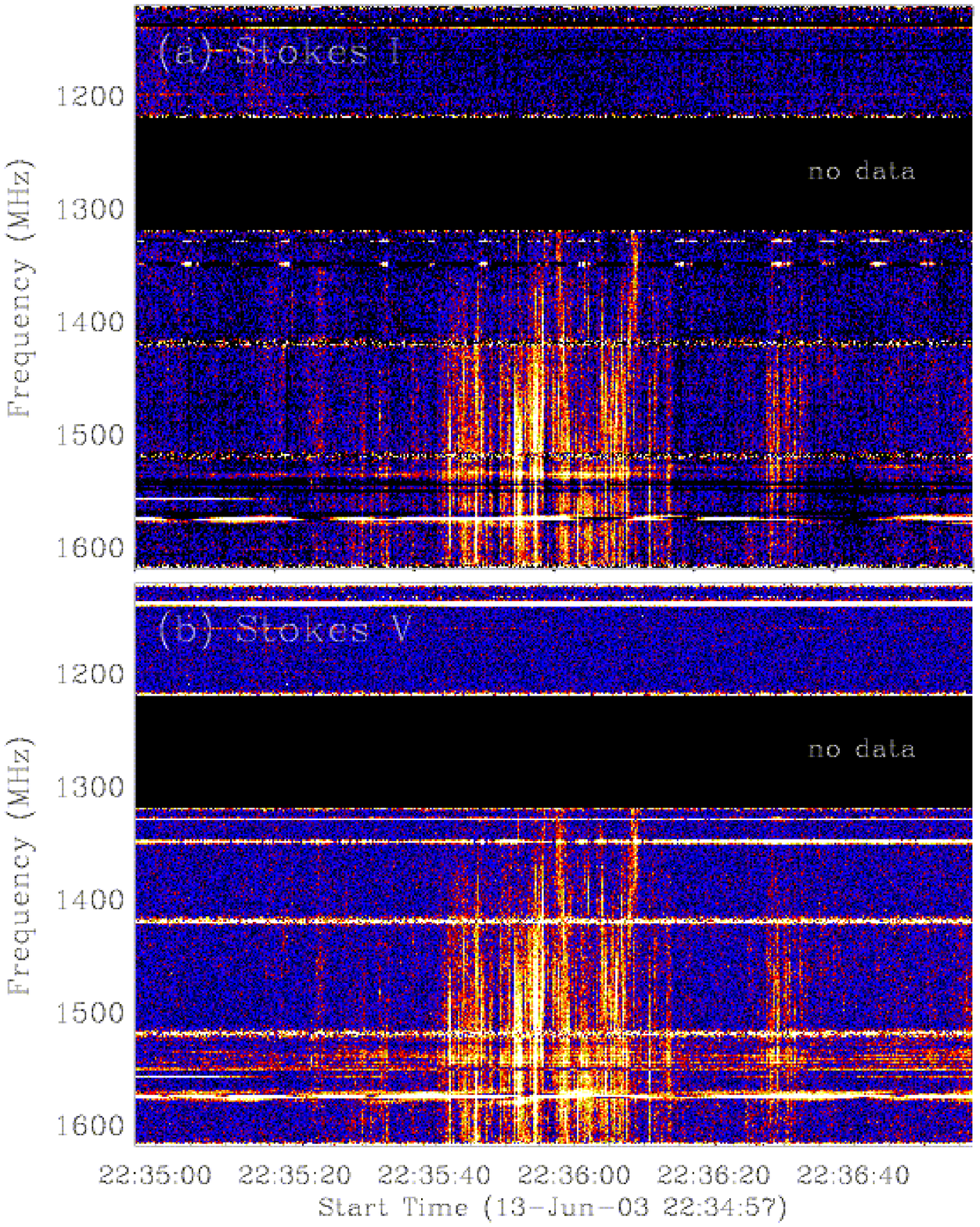}
\figcaption[]{Dynamic spectrum of fast drifting behavior seen
on AD~Leo with the Arecibo Observatory.  Frequency increases down and time
increases to the right.  Frequency coverage extends from 1120--1220 MHz, 1320--1620 MHz; gaps are due to 
RFI excision or poor bandpass response. The data have been rebinned by a factor of 4 in frequency, and 
are illustrated at the full time time resolution of 0.01s. 
{\it (a)} Dynamic spectrum of total intensity
variations.  
{\it (b)} Dynamic spectrum of circular polarization variations.
\label{flare4dynspec}} 
\end{center}
\end{figure}

Figure~\ref{flare4dynspec} displays the dynamic spectrum of the event June 13 in both the Stokes I (total 
flux) and V (circularly polarized flux) parameters.  The radio burst is characterized by the following 
general properties:

\begin{itemize}
\item The burst duration is a few 10s of seconds.
\item The peak flux is $\approx\!200$ mJy.
\item The flux is strongly modulated in time, showing discrete features that have rise times of $\approx\!20$ ms. 
\item The radio burst displays clear substructure in the time-frequency domain.
\item The radio sub-bursts are highly circularly polarized ($>90\%$).
\end{itemize}

\noindent In these respects, the properties of this radio burst are quite similar to those of strong radio 
bursts reported previously \citep{langetal1983,gudeletal1989,bastianetal1990,abada1997a} in this frequency 
band. 
Indeed, noting the fast rise times of discrete sub-bursts, a light travel time 
argument implies source sizes $\l\approx c\Delta t \sim\!9\times 10^8$cm, or 4\% of the stellar 
radius, R$_{\star}$. Using this scale as an estimate of the source size ($R_s < 6000$ km), the observed 
bursts are consistent with brightness temperatures $T_B>4\times 10^{14}$ K.  Such large 
brightness temperatures in AD~Leo's corona require a coherent emission process. 

\begin{figure}[h]
\begin{center}
\includegraphics[scale=0.8,angle=90]{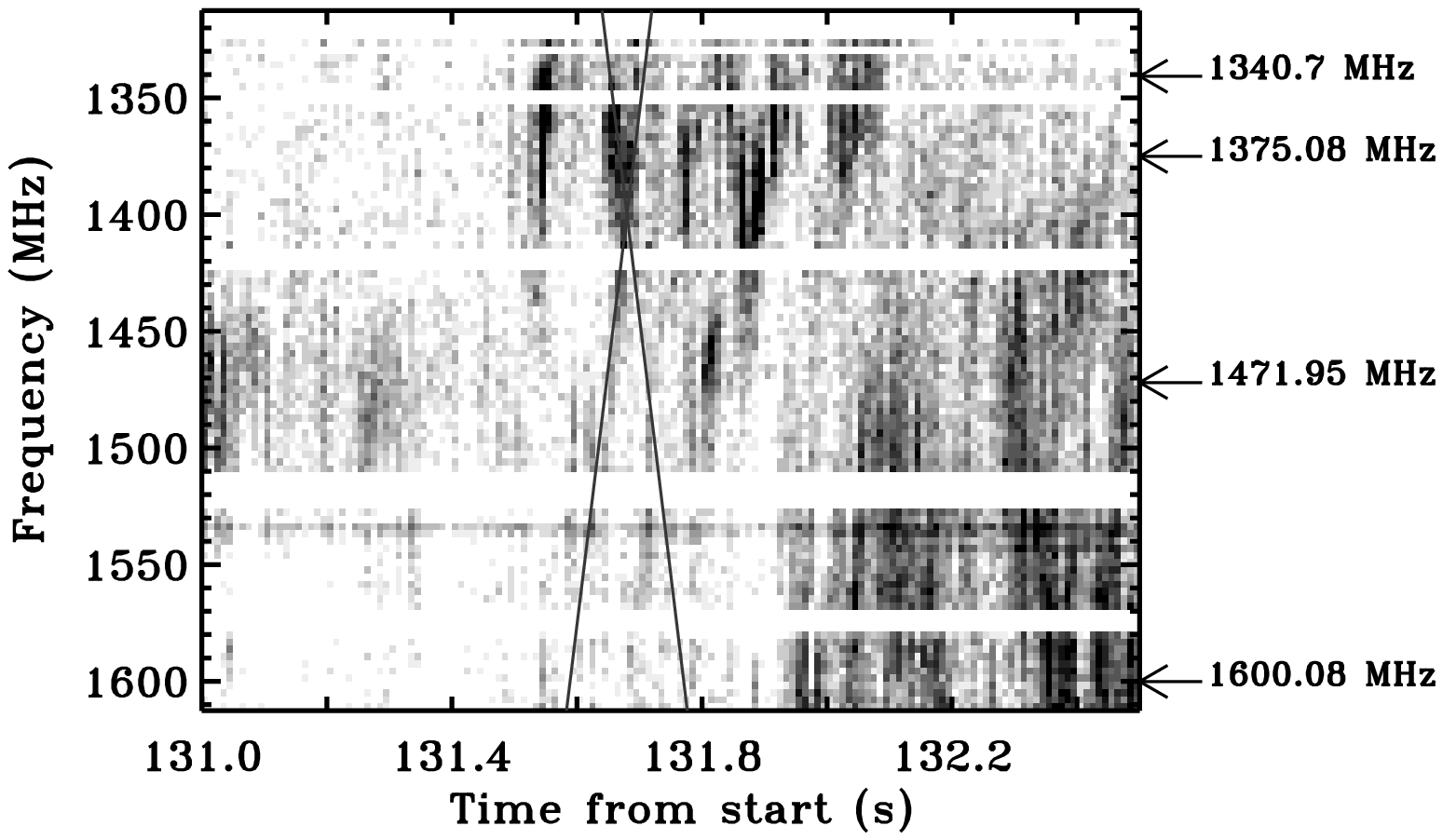}
\includegraphics[scale=0.8,angle=90]{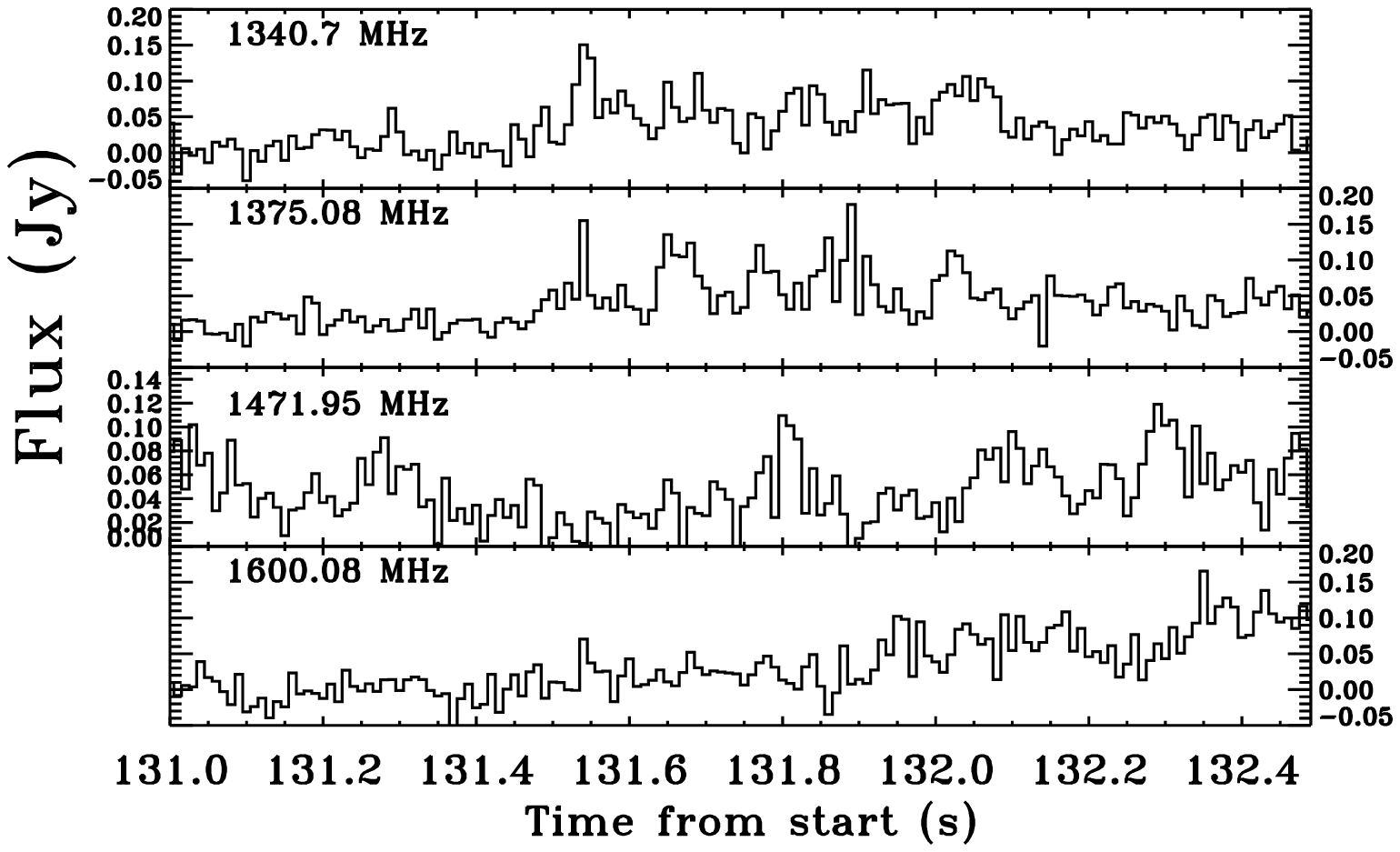}
\figcaption[]{Close up of 1.5 seconds of data in event on 13 June illustrating
sub-bursts with bandwidths typical of the median of the bandwidth distribution.
Top panel displays portion of dynamic spectrum (300 MHz by 1.5 s), bottom panel displays light
curves at four selected frequencies, noted in top panel by arrows.
The black lines in the top panel indicate the characteristic drift rate of 2.2 GHz s$^{-1}$
from a reference time and frequency point of (131.68 s, 1400 MHz), for 
beams directed along and in the opposite sense to the electron density gradient.
\label{fig:closeup}}
\end{center}
\end{figure}

\begin{figure}[h]
\begin{center}
\includegraphics[scale=0.8,angle=90]{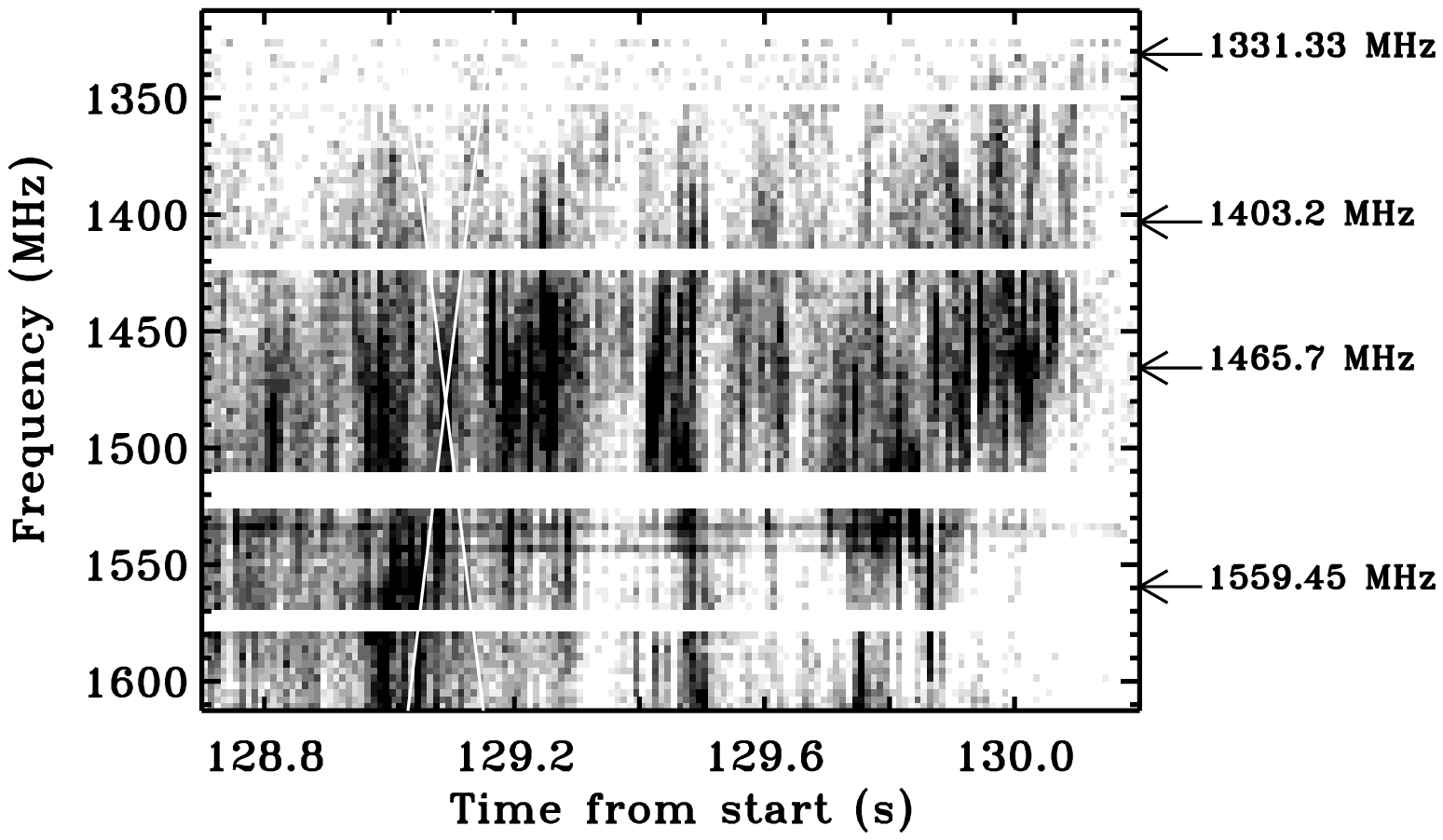}
\includegraphics[scale=0.8,angle=90]{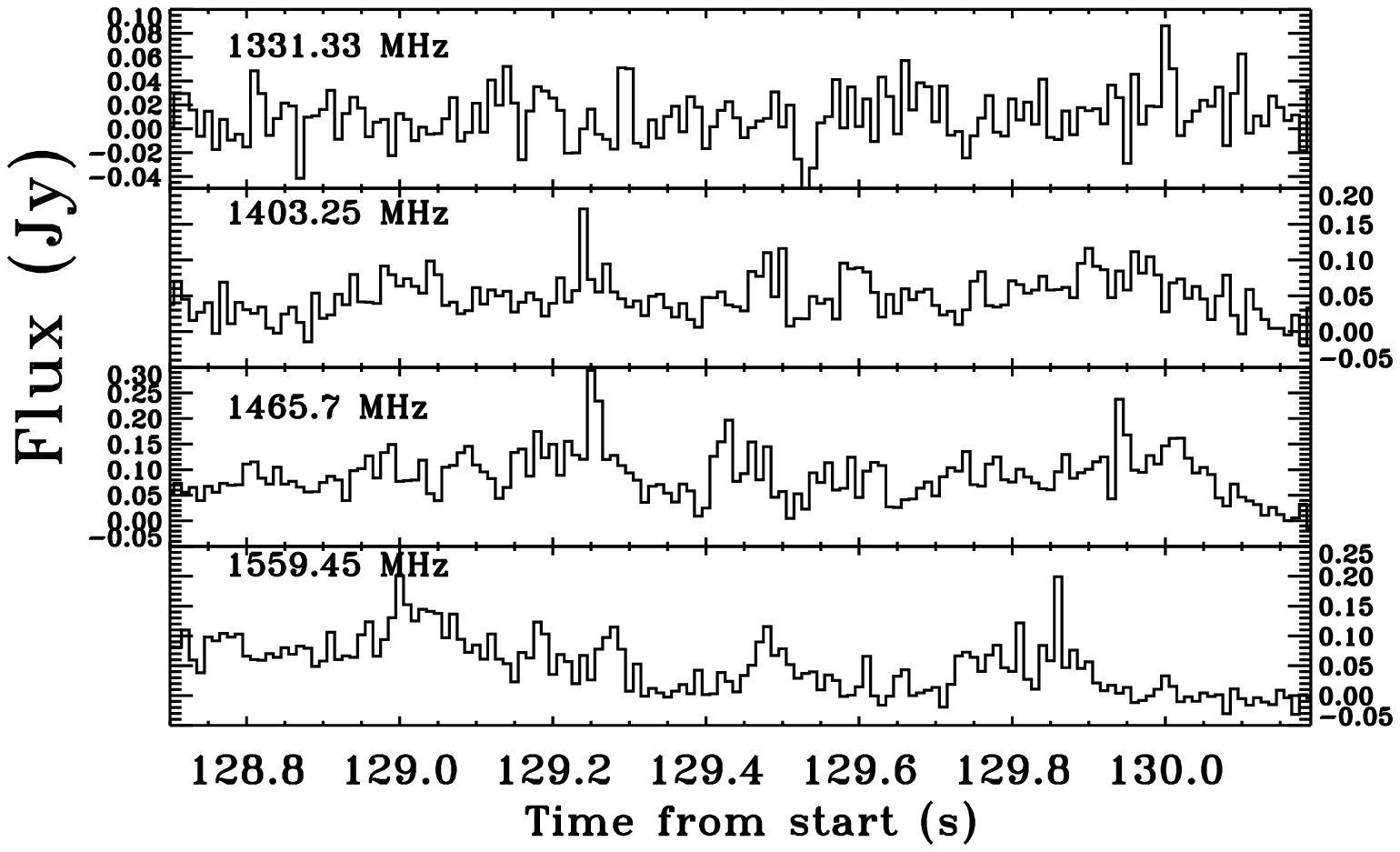}
\figcaption{\label{fig:closeupb}
 Same as for Figure~\ref{fig:closeup} but for sub-bursts which occupy more frequency space.
The white lines in the top panel indicate the characteristic drift rate of 2.2 GHz s$^{-1}$
from a reference time and frequency point of (129.09 s, 1480 MHz),
for beams directed along and in the opposite sense to the electron density gradient.}
\end{center}
\end{figure}

Given the greatly increased bandwidth of these observations, a number of 
new properties of coherent radio bursts on AD~Leo can be identified and analyzed. In particular, 
multitudes of time-resolved and spectrally-resolved structures are present in the spectrum (see 
Figures~\ref{fig:closeup} and ~\ref{fig:closeupb}).  We note that the presence of this fine structure is incompatible with 
diffractive scintillation resulting from propagation of the signal from AD~Leo to the telescope. The 
bandwidth and time scale characteristic of scintillation are expected to be of order several 
GHz and $\sim 10^6$ sec, respectively. We therefore believe the observed fine structure is 
intrinsic to the source. We refer to these fine structures as ``sub-bursts". The start frequency, 
duration, bandwidth, and drift rate of individual sub-bursts can be characterized, offering new 
constraints on the emission mechanism responsible for the observed emission. 

\subsubsection{Methodology}
In order to characterize quantitatively the multitudes of fast-drift radio sub-bursts that comprise this 
event, we employ an algorithm based on \citet{aschwandenbenz1986}, originally used to analyze 
solar radio bursts at decimeter wavelengths. We note that the discrete radio sub-bursts observed on 
AD~Leo are well-resolved in frequency.  Therefore, the dynamic spectrum was binned over 4 adjacent 
frequency channels to improve the signal to noise ratio without loss of spectral information. 
Threshold fluxes were calculated using rms flux values from ``off'' times for each WAPP data channel; 
a 3$\sigma$ flux density threshold of $\approx\!45$ mJy was applied using the binned dynamic spectrum. 
Then, first and second derivatives are computed, and times when the conditions \\
\begin{eqnarray}
\partial S_{ij}/\partial t_{i}>0,\\
 \partial^{2}S_{ij}/\partial t^{2}_{i}<0, \\
 \partial S_{i+1,j}/\partial t_{i+1}<0 
\end{eqnarray}
\noindent where S$_{ij}$ is the flux density at time $i$ and frequency channel $j$, were used to 
identify peaks in each channel.  The time, frequency, and flux
density of each pixel which satisfied the above conditions were recorded.
Individual sub-bursts were identified using an algorithm to group peaks across channels.
This is an IDL procedure adapted from a photometry routine in the IDL
Astronomy User's Library\footnote{Available at http://idlastro.gsfc.nasa.gov/contents.html.}
to
group stars with non-overlapping PSF profiles into distinct
groups.  Instead of using (x,y) as distances, the data points are
(time,frequency) and the distances of any
given point from any other point in time and frequency
are calculated separately.  The critical grouping
describes the distances in time and frequency within which
the peaks must lie to be in the same group. 
We 
explored the probability space of these critical grouping time and frequency values, to determine 
those values in which the probability of having $> n$ peaks occurring at random was sufficiently 
low, $\leq$10$^{-5}$, and found that requiring $n$ to be 5 and restricting t$_{\rm crit} \leq$20 ms, 
$\nu_{\rm crit} \leq$ 20 MHz was sufficient to meet this condition.  Using this procedure, we
identified 313 individual sub-bursts.

Inverse drift rate, start frequency, duration, and bandwidth for the sub-bursts were then
determined. 
We filtered the data for bursts which appeared to continue beyond our frequency limits, 
and required that the flux spectrum of the burst at a given time was Gaussian-like, estimating the 
bandwidth from the FWHM of a third degree polynomial fit to the flux spectrum 
\citep{csillaghy1993}.  The burst duration was calculated as the median FWHM of a parabolic fit
to the temporal variations of all frequency channels in which peaks were identified.
The start frequencies and inverse drift rates were calculated for sub-bursts which
had more than one unique time peak.  
Out of 313 discrete events identified by the above procedure, we determined
bandwidth and duration for 180 sub-bursts,
and inverse drift rate and start frequencies for 121 sub-bursts. 
Figure~\ref{flare4_stat} displays the distributions of these four quantities.

\subsubsection{Rates and Durations}

The median sub-burst duration is 0.03 seconds.
The average 
sub-burst rate is 13 per second during the time when bursting activity is evident, with a maximum 
of $\approx$ 45 s$^{-1}$. Light curves of the bursting behavior in two different bands are 
illustrated in Figures~\ref{fig:closeup} and ~\ref{fig:closeupb}. 
A power spectrum analysis of light curves reveals no evidence 
for periodicities in the data, although previously reported bursts do show 
quasi-periodic oscillations 
in some cases \citep{langwillson1986, bastianetal1990, gudeletal1989, stepanov2001}. 

\subsubsection{Start Frequencies and Bandwidths}

The bursts are confined to frequencies above 1350 MHz, reaching a peak just below a
frequency of 1500 MHz.  
Figure~2 shows that the sub-bursts appear to be fast-drift structures. 
Some structures 
drift to higher frequencies with time, while others drift to lower frequencies with time.

Figure~\ref{flare4_stat} details the distribution of bandwidths.  The median fractional bandwidth 
is 5\%, with individual bursts having measured bandwidths up to 15\%. The minimum and maximum bandwidths 
we are able to determine with our data are $\Delta \nu/\nu$=1\% and 36\%, respectively. 


\subsubsection{Inverse Drift Rates}

\begin{figure}[h]
\begin{center}
\includegraphics[scale=0.7,angle=90]{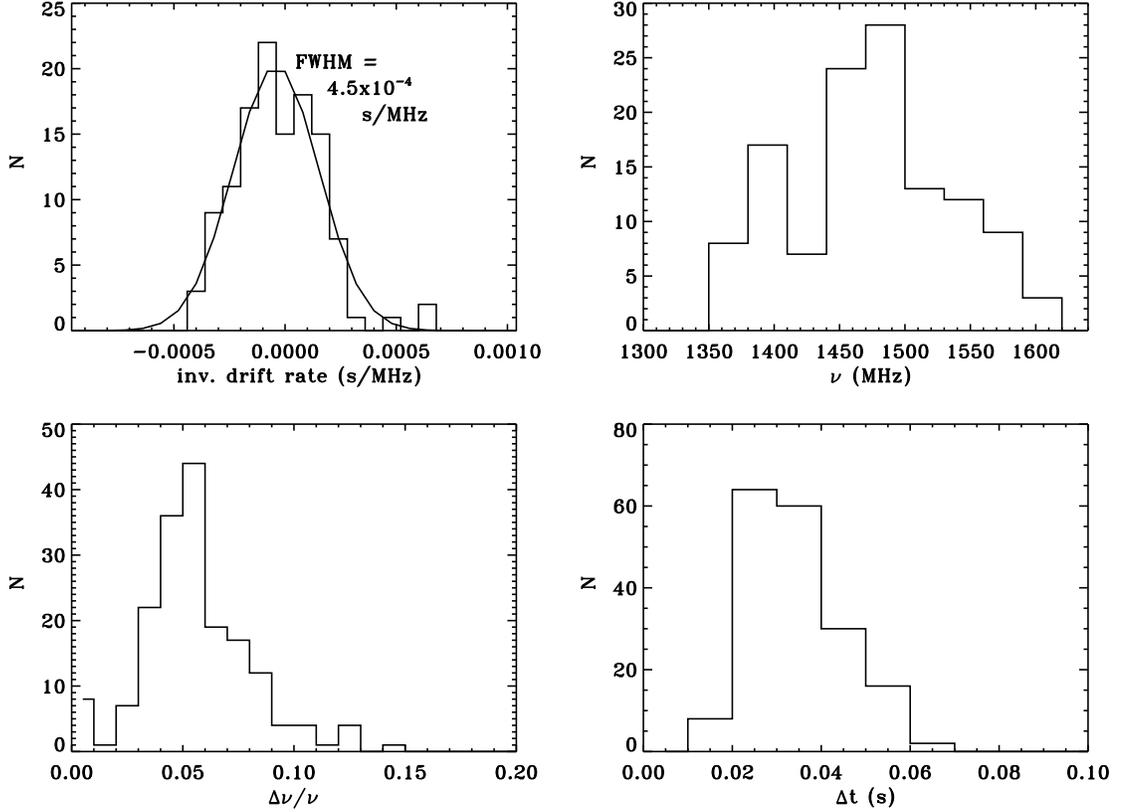}
\figcaption{Distributions of inverse drift rates, start frequencies, bandwidths, 
and durations calculated for bursts seen in the
 1120--1620 MHz spectral region of AD~Leo, during 22:35:00--23:36:40 UT on June 13. 
Top left panel shows inverse drift rate distributions based on time peaks for events with nonzero 
duration and filtered for the flux spectrum shape (see text for details); a Gaussian fit to the 
distribution has a FWHM of 4.5$\times$10$^{-4}$ s/MHz. 
The top right panel displays distribution of start frequencies; this confirms that
most of the sub-bursts occur
at frequencies higher than $\approx$1350 MHz.
Bottom left panel displays bandwidth 
distribution and 
bottom right panel display distribution of durations
for events, subject to a filter; see text
for details.  The median bandwidth is 5\% and the median duration
is 0.03 s.
\label{flare4_stat}}
\end{center}
\end{figure}

Figure~\ref{flare4_stat} displays the inverse drift rate distribution for events with 
nonzero duration;
the observed distribution is symmetric. 
The absolute value of the smallest 
inverse drift rates measurable with our data set arise from a combination of the minimum duration 
and maximum bandwidth, 0.01 s and 487 MHz, respectively, or 2$\times$10$^{-5}$ s MHz$^{-1}$. A 
Gaussian fit to the distribution, illustrated in Figure~\ref{flare4_stat} has a 
FWHM of 4.5$\times$10$^{-4}$ s MHz$^{-1}$, corresponding to a characteristic drift rate of 2.2 GHz s$^{-1}$. 
The distribution of inverse drift rates is incompatible with interstellar dispersion, which would be 
expected to have a value $\approx\!-1.5\times 10^{-6}$ s MHz$^{-1}$ for the dispersion measure
of AD~Leo ($\sim$0.5 pc cm$^{-3}$) at these frequencies \citep[][]{rybickilightman}. 


\subsection{Event on June 14}

\begin{figure}[h]
\begin{center}
\includegraphics[scale=0.6]{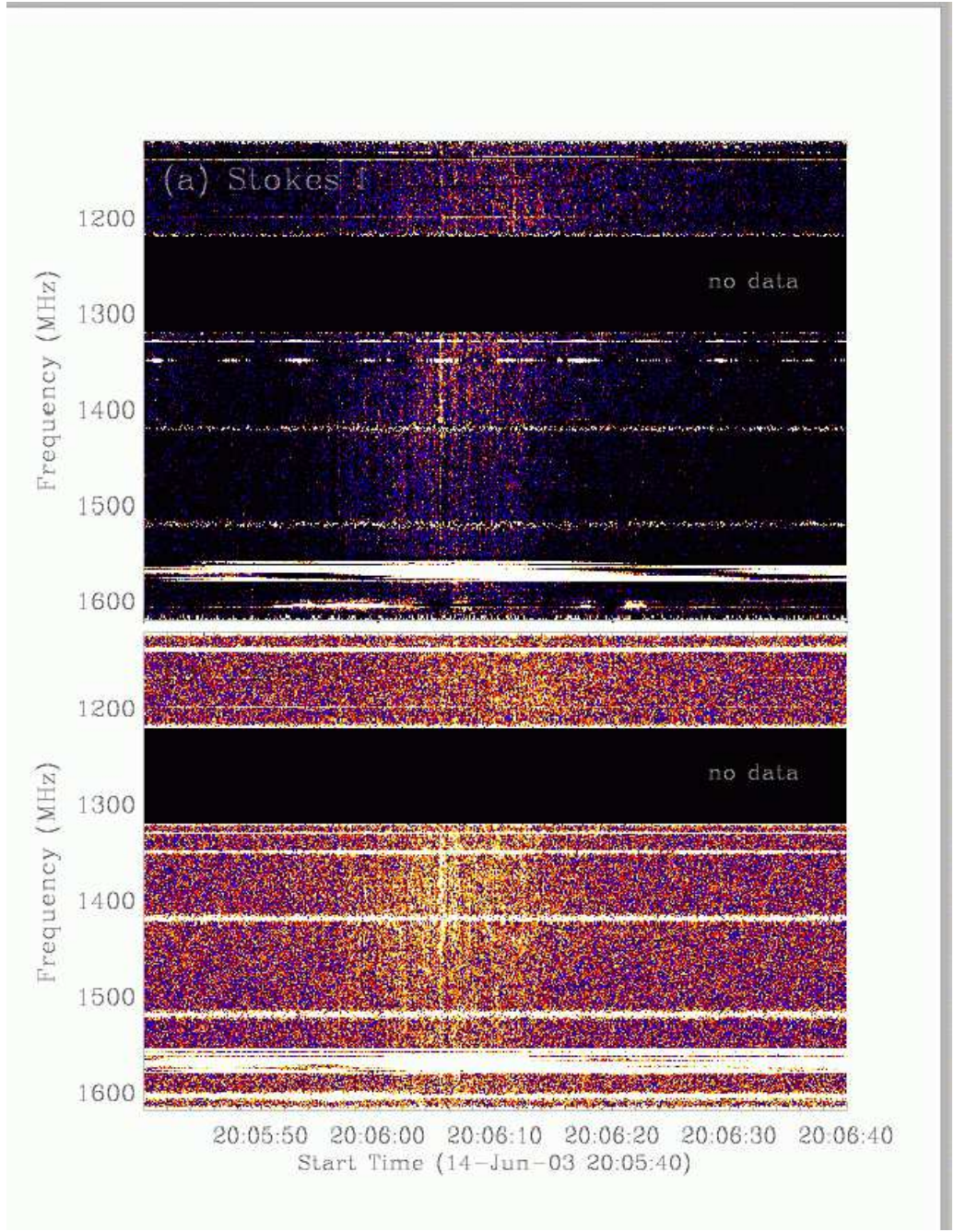}
\figcaption[]{
Dynamic spectrum of slowly drifting behavior seen
on AD~Leo with the Arecibo Observatory.  Frequency increases down and time
increases to the right.  Frequency coverage extends from 1120--1220 MHz, 1320--1589 MHz; gaps are due to
RFI excision or poor bandpass response. 
{\it (a)} Dynamic spectrum of total intensity
variations.
{\it (b)} Dynamic spectrum of circular polarization variations.
\label{flare5dynspec}}
\end{center}
\end{figure}

The dynamic spectrum of the event on June 14 in Stokes I and V is displayed in Figure~\ref{flare5dynspec}. 
While similar in duration to the June 13 event, it otherwise differs significantly from it. The 
June 14 emission is characterized by a smooth variation with time rather than the spiky emission seen 
on June 13. 
By eye the event appears to contain sub-structure, and we explored the possible
existence of quasi-periodic oscillations in the data.  Although there were peaks in the periodograms
at 0.5 and 40 Hz
for each rebinned channel, 
these periods appeared in the data directly preceding and following 
the event as well, casting doubt on the reality of finer temporal scale structures in the event.

\begin{figure}[h]
\begin{center}
\includegraphics[scale=0.7]{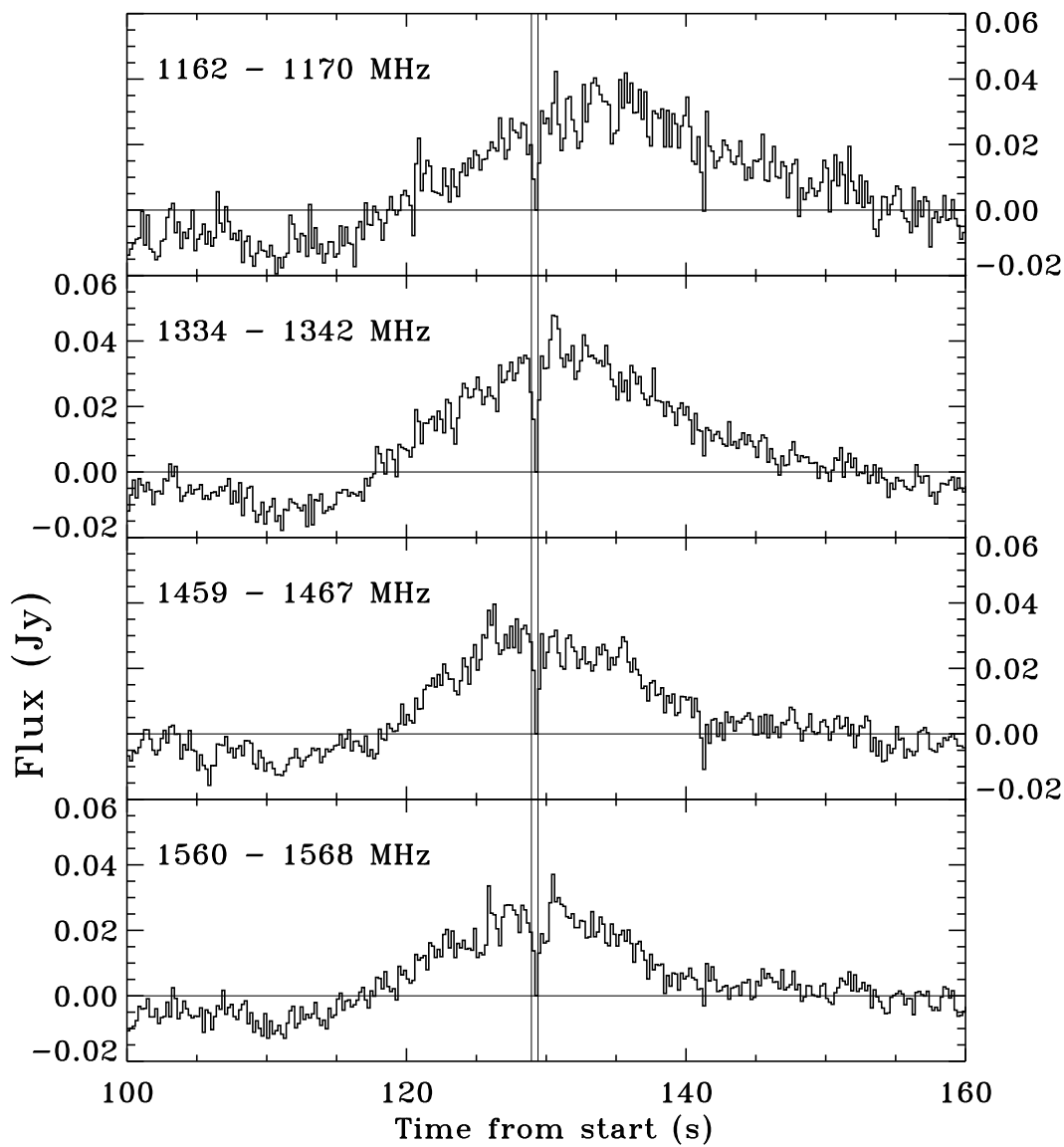}
\figcaption{ Light curves of event on June 14 at four different frequency ranges.
Data have been binned by a factor of 10 in frequency and 20 in time.
The data dropout at $\sim$129 seconds at all frequencies is due to the appearance of a periodic radar
signal causing RFI; this data has been discarded.
This event is characterized by a smooth, Gaussian-like modulation and the drift to lower frequencies at
later times is evident.
\label{fig:flare5lc} }
\end{center}
\end{figure}

\begin{figure}[h]
\begin{center}
\includegraphics[scale=0.7,angle=90]{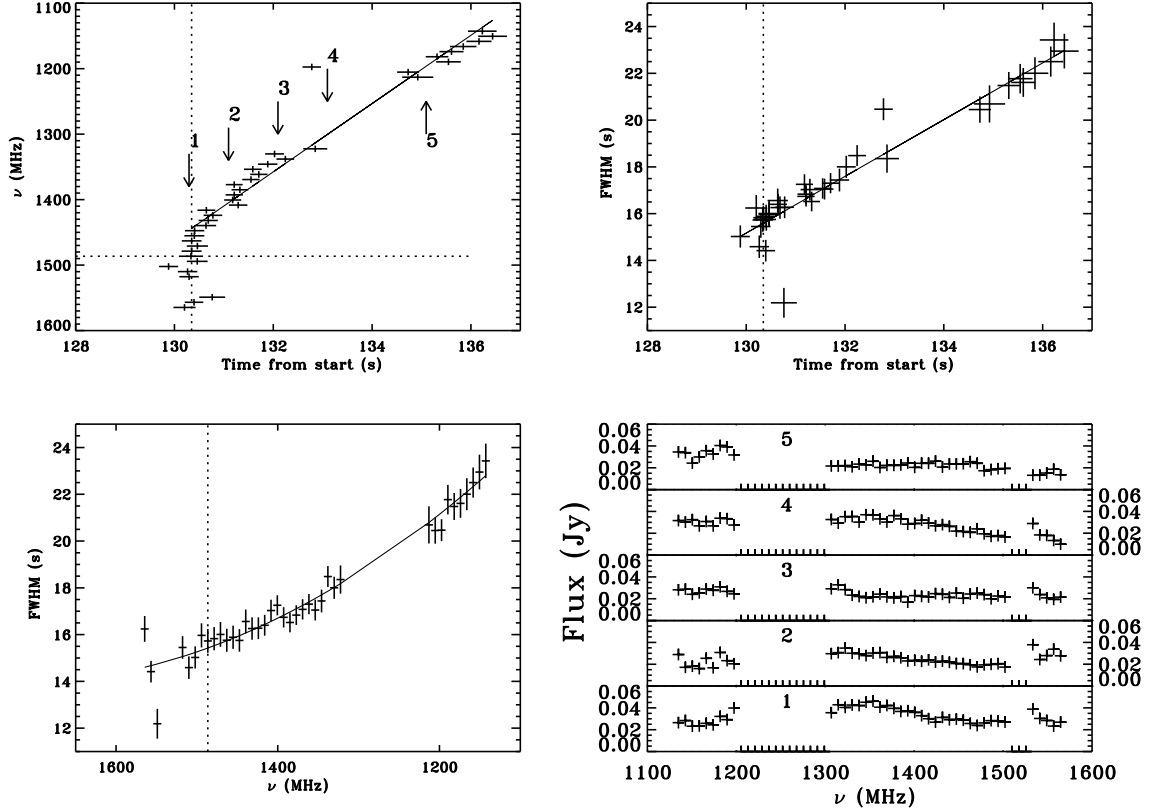}
\figcaption[]{Variation of Gaussian parameters during event on 14 June 2003.  Top left panel shows the central time of the Gaussian fit for each frequency bin; the linear fit to $\nu\leq 1490$
MHz is $-$52 MHz s$^{-1}$.
Numbers indicate times at which flux spectra are illustrated in bottom right panel.
Dashed vertical and horizontal lines indicate the approximate time and frequency, 
respectively, of the burst start.
Top right panel shows increase of Gaussian FWHM with time; the slope of the linear fit
is 1.2 FWHM (s) s$^{-1}$. The vertical dashed line indicates the approximate time
at which the burst started, as deduced from the upper left panel.
Bottom left panel displays increase of burst duration towards smaller
frequencies; the solid line is a second degree polynomial fit to the data. 
The vertical dashed line indicates the approximate frequency at which the burst started.
Bottom right panel displays the flux versus frequency at several points during the burst; 
numbers refer to times denoted in top left panel. 
\label{flare5params}}
\end{center}
\end{figure}

Light curves (Figure~\ref{fig:flare5lc}) show 
that the time variation of the emission has an approximately Gaussian variation. 
Both the duration of the emission and the time of peak flux evolve towards lower frequencies. 
The data were 
averaged in frequency by a factor of 10 and in time by a factor of 20 
to ameliorate the statistics, and were then fit 
to a Gaussian time profile in each frequency bin. Figure~\ref{flare5params} displays the variation 
of the Gaussian parameters (center, width, peak). The peak times of the Gaussians drift 
with frequency 
at a rate of $-52$ MHz/s from 1120--1490 MHz, a drift rate that is far smaller than those characteristic 
of the sub-bursts observed on June 13. The event apparently extends to both lower and higher frequencies. 
The lower limit on the bandwidth of the emission is therefore $\Delta\nu/\nu>36\%$. The duration 
of the event decreases with frequency.  As was the case for the June 13 event, the degree of 
circular polarization is high ($>50$\%), but less well constrained than the June 13 event as a result of the poorer signal to noise ratio.

The lack of temporal and spectral fine structures prevents us from using such features to 
constrain the source size. Presuming the source is comparable in scale to the star itself 
($l_S\approx R_\star\approx 0.5R_\odot$), a brightness temperature of $\approx\!2\times 10^{11}$ K is 
obtained. This might suggest that an incoherent emission mechanism (e.g., synchrotron radiation) is 
feasible. However, the high degree of circular polarization indicates otherwise. We conclude that, 
like the June 13 event, the June 14 event is likely due to a coherent emission mechanism. If so, the 
source scale $l_S$ may be considerably smaller than the size of the star: 
$l_S \approx 5\times 10^{10} {\rm T_{B10}^{-1/2}}$ cm, where ${\rm T_{B10}}$ is the brightness 
temperature in units of $10^{10}$ K.  

\section{Discussion}

We have concluded that, given the extremely high brightness temperatures and the high 
degree of circular polarization observed, a coherent emission mechanism or mechanisms is/are 
responsible for the radio bursts observed on AD~Leo on 2003 June 13 and 14. We now discuss each 
event in turn.

\subsection{June 13 Event}

It is useful to draw parallels between the properties of the June 13 radio burst observed on AD~Leo 
and radio bursts observed in the solar corona, recognizing that such parallels must not be given 
undue emphasis. A summary of different types of solar radio bursts is given in Table~1 of 
\citet{dulk1985}. More recently, work on decimeter- wavelength and microwave solar bursts has been 
summarized by \citet{bbg1998}.  A striking feature of the June 13 event is the presence of 
multitudes of fast drift sub-bursts. This might suggest that a phenomenon analogous to solar decimetric 
type III bursts (type IIIdm) or solar spike bursts may be relevant.

Like meter wavelength type III radio bursts, type IIIdm bursts are fast-drift bursts that occur as a 
result of the nonlinear conversion of plasma waves (Langmuir waves) to electromagnetic radiation 
at the local electron plasma frequency $\nu_{pe}=\sqrt{e^2 n_e/\pi m_e}\approx 9n_e^{1/2}$ kHz 
or its harmonic $2\nu_{pe}$. The Langmuir waves are believed to be excited by the passage of a 
suprathermal electron beam propagating at a speed of $\sim 0.1-0.3$c.  If the beam propagates from 
lower (denser) to greater (more rarefied) heights in the corona, the radio emission drifts from high 
to low frequencies. If the beam propagates from greater to lower heights, the radio emission drifts 
with time in the opposite sense. In some respects, type IIIdm may be regarded as the high frequency 
extension of classical type III radio bursts. However, 
for metric type III bursts the 
vast majority drift from high to low frequencies, indicating a beam exciter that 
propagates from high to low densities in the corona. 
The same is not true for type IIIdm bursts. 
In particular, the majority of type IIIdm bursts drift with frequency in the opposite sense to 
that of metric type IIIs, indicating most are due to electron beams propagating from lower to 
higher densities. 

Closer examination of the properties of the sub-bursts determined in \S3.1 suggests that the June 13 
event bears closer resemblance to solar decimetric spike bursts.
Decimetric spike bursts have generated interest since their discovery \citep{droege1977}
as phenomena possibly closely linked with magnetic energy release in the solar corona \citep{benz1986}. 
They differ from type III dm bursts in several critical respects. First, their instantaneous 
bandwidths typically are an order of magnitude less than those of type IIIdm bursts. Second, their 
durations are also roughly an order of magnitude smaller than those of type IIIdm bursts. 
Indeed, the decay times of the decimeter spike bursts may be consistent with electron-ion collisional 
damping \citep{gudelbenz1990}, unlike those of type III and type IIIdm bursts. Third, again in 
contrast to type IIIdm bursts, the degree of circular polarization of spike bursts tends to be 
high, at least for those observed on the disk of the Sun where propagation effects are expected to be 
small \citep{gudelzlobec1991}. Unlike type III or type IIIdm bursts, the distribution of frequency 
drifts shows no strong preference for predominantly positive or negative drifts. A study by 
\citet{gudelbenz1990}
demonstrated that the distribution of {\sl inverse} drift rates of solar decimetric spikes 
is roughly symmetric at 362 and 770 MHz. These were interpreted as a nearly isotropic 
distribution of exciter directions within the context of plasma radiation.  

To a striking degree, the June 13 event is similar in its properties to solar decimetric spike bursts, 
an exception possibly being the durations of the discrete bursts, which appear to be a few times 
longer than solar spike bursts at the same frequency, a point to which we return below. While 
plasma radiation produced by the nonlinear conversion of Langmuir waves excited by the passage 
of an electron beam is the accepted emission mechanism for type III and type IIIdm bursts on the 
Sun, the emission mechanism for decimetric spike bursts is more controversial. One possibility 
is the electron cyclotron maser (ECM) mechanism, which operates at the fundamental or harmonic 
of the local electron gyrofrequency $\Omega_{Be}=eB/2\pi mc =2.8B$ MHz 
when $0.3<\omega_{pe}/\Omega_{Be}<1$ \citep[see, e.g.,][for a review]{fleishman1998}.
The other possibility is a plasma radiation process. Neither mechanism has been shown 
unambiguously to be the mechanism responsible for solar decimetric spike bursts. Likewise, 
neither mechanism can be firmly identified as that responsible for the spike bursts on AD~Leo 
although, for reasons we now outline, a plasma radiation mechanism is favored.


The primary reason we favor plasma radiation over the ECM is that in the hot, dense corona of 
AD~Leo, radiation generated by the ECM at the first or second harmonic of $\Omega_{Be}$ can be 
readily gyroresonantly absorbed at the second or third harmonic layers, respectively. Although 
the situation may be avoided or mitigated \citep[see][and references therein]{fleishman1998} in the solar corona, the
problem is exacerbated in the corona of AD~Leo, which is significantly hotter than the Sun's. 
Whereas the bulk of solar X-ray-emitting thermal plasma in quiescence is 1--2 MK, AD~Leo's corona 
is dominated by 6--20 MK soft X-ray-emitting plasma. Since the gyroresonance 
absorption coefficient increases as $\kappa_{gr}\sim T^{s-1}$, where $s=2,3$ is the harmonic of the 
absorbing layer, the absorption is significantly greater in the corona of AD~Leo than it is in the 
Sun's corona.  In contrast to ECM radiation, plasma radiation is most susceptible to free-free absorption. 
Unlike gyroresonance absorption, however, the efficacy of free-free absorption decreases 
with increasing plasma temperature as $\kappa_{ff}\propto T^{-3/2}$. Hence, the higher temperature of AD~Leo's corona 
favors the escape of radiation generated via a plasma radiation mechanism over that generated by the 
ECM mechanism if $\omega_{pe}>\Omega_{Be}$. For fundamental plasma radiation, this suggests $B<500$ G, 
a condition easily and widely met in the corona of AD~Leo. Moreover, \citet{zaitsev2000} have 
argued that plasma radiation mechanisms operate significantly more efficiently in the hot corona of 
late-type stars.  


Assuming, therefore, that the burst emission on AD~Leo is due to fundamental plasma radiation, the 
symmetry of the inverse drift rate distribution implies an isotropic distribution of exciters 
with respect to the electron density gradients.  If a given exciter is a beam propagating through 
AD~Leo's corona, the drift rates imply \citep[e.g.,][]{benz2002}\\
\begin{equation}
\frac{d\nu}{dt} = \frac{\partial \nu}{\partial n_{e}} \frac{\partial n_{e}}{\partial h} \cos\phi\frac{\partial s}{\partial t},
\end{equation}
where $n_{e}$ is the ambient electron density, $h$ is the vertical height, 
$\partial h = \partial s\cos\phi$, $\cos\phi$ is the angle between the beam direction and the 
vertical, and $\partial s/\partial t=v$ is the propagation speed. Approximating the density 
variation as an exponential with a scale height $H_{n}$, this equation can be 
rewritten $\dot\nu \sim -(\nu v\cos\phi)/(2H_{n})$ and so $v\sim -2H_n\dot\nu/\nu\cos\phi$. 
Taking the FWHM inverse drift rate of $4.5\times 10^{-4}$ s MHz$^{-1}$, 
$\dot\nu=2.2$ GHz s$^{-1}$, $v\sim4H_n$ for $\phi\approx 45$ degrees. 
Constraining $v$ to be $0.1c-0.5c$ suggests a scale height of order $1-4\times 10^9$ cm.
The isothermal scale height for a 10 MK plasma around a star of mass $\sim$0.3 M$_{\odot}$
and radius $\sim$0.3R$_{\odot}$ is about a factor of ten larger than this scale height,
suggesting that the scale height deduced from the drift rate may indicate an
inhomogeneity size scale in AD~Leo's corona within which the sub-burst exciters
are propagating. 
Interestingly, if we constrain the event duration ($\tau \sim $0.03 s) to be of order the Coulomb 
deflection time, the temperature in the source can in turn be constrained: \\
\begin{equation}
T\sim 8000 \nu_6^{4/3} \tau^{2/3}\approx 13{\rm MK} 
\end{equation}
where $\nu_6$ is the observed frequency in MHz. This value is comparable with coronal temperatures 
inferred from SXR observations.

\subsection{Event of June 14}

The second event is markedly different from the June 13 event. It is characterized by a smoothly 
varying continuum in time and frequency. As shown in \S3.2, the time variation of the emission 
at a given frequency is well-approximated by a Gaussian profile. The Gaussian width and the 
time of maximum progressively increase with decreasing frequency. In contrast to the 
fast-drift bursts in the June 13 event, which had characteristic drift rates of 2.2 GHz s$^{-1}$, the 
drift rate of the June 14 event is $-52$ MHz s$^{-1}$. The drift of the event could be due to a 
disturbance propagating through AD~Leo's corona.  Under this interpretation, the observed 
drift rate value of $-52$ MHz s$^{-1}$ implies that the disturbance is propagating outward; 
taking $\nu=1490$ MHz for the approximate starting frequency, at coronal temperatures 
($T=10^{6},10^{7}$K), the observables constrain the propagation speed to be (1200, 1.2$\times$10$^{4}$) km s$^{-1}$, 
assuming that it is fundamental plasma emission. 
Associating the observed frequencies with fundamental plasma emission produces a constraint on
the ambient electron density, $n_{e} \leq$3$\times$10$^{10}$ cm$^{-3}$, which is consistent with
measurements of cool SXR-emitting coronal material \citep[$n_{e} \leq$3$\times$10$^{10}$ cm$^{-3}$
at T$\sim$2MK;][]{vdb2003}.
These speeds are consistent with the Alfv\'{e}n speed 
in a medium with $B\leq$ 100 (1000) G, $n_{e}\leq 3\times10^{10}$ cm$^{-3}$ for 1200 (1.2$\times$10$^{4}$) km s$^{-1}$, 
respectively.  
The maximum duration (FWHM) is $\sim$23s, which implies that the extent of the 
atmosphere through which the disturbance propagated is 0.1 (1) R$_{\star}$, for $v=$1200 (1.2$\times$10$^{4}$) 
km s$^{-1}$, respectively.  Such a source size, coupled with the estimates of the maximum flux 
density, constrain the brightness temperature to be 
6$\times$10$^{11}$--6$\times$10$^{13}$K.

\section{Summary and Conclusions}


We have presented wideband dynamic spectra of two radio bursts observed on the dMe flare star AD~Leo. 
The observed properties of the two bursts indicate that a coherent emission mechanism is responsible. 
A plasma radiation mechanism is favored. The first event, on 2003 June 13, contained multitudes of 
fast-drift bursts that were similar in their properties to solar decimetric spike bursts. 
In particular, the distribution of their bandwidths, inverse drift rates, and the high degree of 
circular polarization were qualitatively and quantitatively similar to solar spike bursts. The 
distribution of durations were a factor of $\sim 3$ greater than their solar analogs. The 
observations are consistent with the idea that the spike bursts are due to fundamental 
plasma radiation driven by electron beam exciters. The burst durations are consistent with the 
collisional decay time. We attribute the fact that they are longer-lived than their solar analogs 
to the greater temperature of the coronal medium on AD~Leo. The second event, on 2003 June 14, was 
different in character from the first event, being characterized by a smoothly varying continuum 
and frequency drift rate significantly smaller than the fast-drift bursts on June 13. The emission 
mechanism responsible for the burst is less certain, but if it is plasma radiation, the observed 
drift rate is consistent with a disturbance propagating in the stellar corona with a speed comparable 
to the local Alfv\'{e}n speed.



We conclude by pointing out that the upgrades to the Arecibo Observatory and the Green Bank Telescope 
now open up wide regions of frequency space which can be observed simultaneously
at high time resolution.  These 
advances in technology allow for dynamic spectrum analysis of bursting phenomena on stars using 
techniques used for decades on the Sun.  Such instrumental advances allow an investigation of 
plasma physics in stellar environments on the smallest spatial and temporal scales yet obtained.  
These investigations are crucial to establishing the extent to which solar phenomena can be applied 
in extreme stellar environments.

\acknowledgements{We thank Phil Perillat and Avinash Deshpande at NAIC for their outstanding assistance 
in making this observational program a success. 
We thank the anonymous referee, Manuel G\"{u}del, for his enthusiastic reading of the manuscript.
This paper represents the results of program A1730 at Arecibo Observatory.}


\clearpage

\end{document}